\documentclass[10pt]{article}
\usepackage{fullpage,citesort,epsfig,psfrag,graphics,amsbsy}
\usepackage[normal,small]{caption}
\newcommand{\beq}{\begin{equation}}
\newcommand{\eeq}{\end{equation}}
\newcommand{\bea}{\begin{eqnarray}}
\newcommand{\eea}{\end{eqnarray}}

\newcommand{\be}{\begin{equation}}
\newcommand{\ee}{\end{equation}}
\newcommand{\bq}{\begin{eqnarray}}
\newcommand{\eq}{\end{eqnarray}}

\def\math{\mathsurround=0pt }
\def\leftrightarrowfill{$\math \mathord\gets \mkern-6mu 
 \cleaders\hbox{$\mkern-2mu \mathord- \mkern-2mu$}\hfill
 \mkern-6mu \mathord\to$}
\def\overleftrightarrow#1{\vbox{\ialign{##\crcr
     \leftrightarrowfill\crcr\noalign{\kern-1pt\nointerlineskip}
     $\hfil\displaystyle{#1}\hfil$\crcr}}}

\def\dalemb#1#2{{\vbox{\hrule height .#2pt
        \hbox{\vrule width.#2pt height#1pt \kern#1pt
                \vrule width.#2pt}
        \hrule height.#2pt}}}

\def\0{{(0)}}
\def\9{{(9)}}
\def\8{{(8)}}
\def\7{{(7)}}
\def\6{{(6)}}
\def\5{{(5)}}
\def\4{{(4)}}
\def\3{{(3)}}
\def\2{{(2)}}
\def\1{{(1)}}

\begin{document}
\setlength{\captionmargin}{20pt}

\begin{center}
\begin{large}
{\bf String/Flux Tube Duality}
\footnote{Supported in part by the Department
of Energy under Grant No. DE-FG02-97ER-41029.} 
\end{large}
\vskip5pt
{\large 
Charles B. Thorn\footnote{E-mail  address: {\tt thorn@phys.ufl.edu}}
}
\vskip5pt
{\it %Institute for Fundamental Theory\\
Department of Physics, University of Florida,
Gainesville FL 32611}
\end{center}
\begin{abstract}
I describe Field/String duality as applied to the response 
of gauge fields to
separated quark and antiquark sources. This is a talk contributed
to the conference {\it Quark Confinement and the Hadron Spectrum
VII},
Ponta Delgada, Azores, 2-7 September 2006.
\end{abstract}
\section{Introduction}
The physics of quark confinement is generally believed to
involve two conjectured properties of quantum Yang-Mills gauge theory: 
(1) that
there is  a mass gap $m_G$ (the lightest glueball mass)
and (2) that the gauge field responds to a fixed $Q$ source separated from
a fixed ${\bar Q}$ source by a distance $L$ 
by forming a gluonic flux tube (or gluon chain) between $Q$
and ${\bar Q}$ with energy $U(L)\simeq T_0 L$ for large $L$.

Although both of these facets of quark
confinement are firmly established numerically, an
analytic understanding of either is so far unattained. 
Indeed just proving the mass gap 
is one of the Clay Institute millennium problems.
My message here is that a less daunting path to such an
analytic understanding may be the
reformulation of Yang-Mills as a String Theory (Field/String Duality)
\cite{maldacena,adscft,bardakcit}.
This reformulation can proceed without solving the theory
or even without establishing a mass gap.
It might well provide both a setting, in which the physics
of confinement can be understood by using
string variables to construct a tractable model of the gluonic 
flux tube (gluon chain), and a vehicle for 
a self-consistent determination of a mass gap.

The AdS/CFT correspondence shows that a string reformulation
of QFT is quite independent of confinement and also
of the existence of a mass gap. 
Indeed, the best understood case of Field/String duality
is the equivalence of ${\cal N}=4$ supersymmetric SU(N) Yang Mills
to IIB superstring theory on
AdS$_5\times$S$_5$ \cite{maldacena}.
In this case the finiteness of the ${\cal N}=4$ theory
implies conformal invariance which in turn implies
a vanishing mass gap and zero string tension. On the string side these 
features are consequences of the curved AdS background.
The string interpretation is particularly transparent
when $N\to\infty$ because $g_{\rm string}\sim1/N$, so it is a limit
in which strings do not break or interact.

From this point of view string theory offers something much
more tangible to theoretical physics than a nebulous and
quasi-religious ``theory of everything''. I 
believe that it provides a practical theoretical framework
for resolving some of the most intriguing 
conundrums of quantum field theory. In this regard, I
offer a new definition of string theory by way of an analogy:
\begin{center}
\vskip5pt
\noindent
String Theory : $\sum$ (Planar Diagrams) :: 
Bethe-Salpeter Equation :$\sum$ (Ladder Diagrams) 
\vskip5pt
\end{center}

Just as the sum of ladder diagrams
gives a zeroth order Bethe-Salpeter equation 
so does the sum of planar diagrams give a zeroth order (noninteracting)
string theory. In both cases the full QFT can be regained by
systematic corrections. For string theory the nonplanar corrections
are neatly handled via 't Hooft's $1/N$ expansion \cite{thooftlargen},
since the planar approximation is exact in 
the large $N$ limit with $\lambda=N\alpha_s/\pi$ fixed.

In this talk I shall explain how stringy features appear 
in the conformally invariant ${\cal N}=4$ case. Though this theory 
lacks a mass gap and quark confinement, it
nonetheless produces a stringy gluonic flux tube between separated
color sources. Moreover, we can easily reach interesting conclusions
about this flux tube's physical properties, especially in the
strong 't Hooft coupling limit when the
string can be treated semi-classically. This limit makes sense 
for ${\cal N}=4$ because the coupling does not depend on the scale,
and is thus a free parameter.
In a string theory formulation of pure Yang-Mills, such a semiclassical limit
is not possible because there is no tunable coupling. 
Nonetheless, the conformal case is an important
example because it shows a limit in which the planar sum can actually
be done. Although there should be a strikingly different outcome
for the sum of planar diagrams in QCD, 
the technical difficulties in the two problems are quite
comparable. Solving one should teach us a great deal about solving the
other.
\section{Separated $Q{\bar Q}$ Sources}
At $N=\infty$ the response of the ${\cal N}=4$ theory 
to separated static color sources is very interesting: a flux tube
forms with an excitation spectrum that becomes string-like in the
limit of strong 't Hooft coupling $\lambda\to\infty$. A 
convenient probe that reveals this excitation spectrum is the
expectation of a rectangular
$L\times T$ Wilson Loop, $\langle W(L,T)\rangle\sim \sum_n w_n\exp(-TE_n(L))$
as $T\to\infty$. The $L$ dependence of the ground state energy tests
the presence of confinement, a test the ${\cal N}=4$ theory fails because it
is conformal $E_G(L)\sim -c/L$ at large $L$. But the excitation spectrum
$E_n(L)$ at fixed $L$ can look stringy and does at strong 't Hooft coupling.  

\begin{figure}[ht]
\psfrag{'T'}{$T$}
\psfrag{'L'}{$L$}
\psfrag{'sim'}{}
\begin{center}
\includegraphics[width=5in]{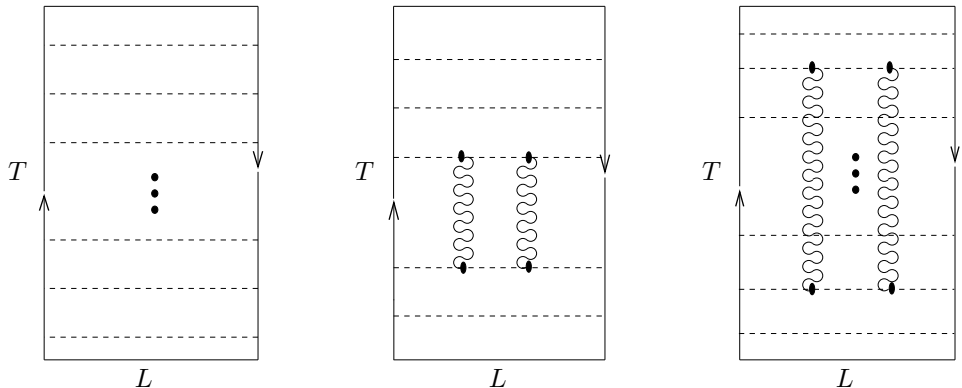}
\caption{Wilson loop with only Coulomb exchange (left), 
a planar radiative correction (middle), 
and a nonplanar correction (right).}
\label{coulomb}
\end{center}
\end{figure}

First we consider the weak coupling limit of this $Q{\bar Q}$
system, $\lambda \ll1$. In this limit in
Coulomb gauge the planar approximation to the Wilson loop, in pure
Yang-Mills theory, is
given by multiple instantaneous Coulomb exchanges (See Fig. \ref{coulomb}).
The Wilson loop ensures $Q{\bar Q}$ are in a singlet. We can 
therefore read off the singlet energy as the coefficient of $-T$ in
the exponential behavior at large $T$. In the
case of the sum of diagrams with only Coulomb exchange,
this shows $E_{singlet}=-\pi\lambda/2L$,
and this is the only eigenstate that couples.
The resolvent shows only a single pole
$$R_0(E,L)\equiv\int_0^\infty dT e^{ET}W_0 \sim 
{\rho_0^\prime\over -E-\pi\lambda/2L}$$ 
Note that in the ${\cal N}=4$
conformal theory the Wilson loop is modified to include coupling to the
scalar fields. This (1) doubles the effect of Coulomb exchange, so
we have to understand 
$\lambda\to 2\lambda$ in this formula, and (2) introduces a continuum for $E>0$
at lowest order, because
the scalar propagator is not instantaneous.
A very interesting feature of this continuum as well as that due
to planar radiative corrections
(e. g. the middle diagram of Fig.~\ref{coulomb}) is that, 
since planarity forbids 
gluon exchanges between parts of the diagram
separated by propagating transverse gluons,
there is a gap between the discrete ground state and the continuum,
summarized by the following form for the planar resolvent 
(now for the ${\cal N}=4$ theory) 
$$R_{\rm planar}(E,L)\sim 
{\rho_0\over -E-\pi[\lambda+ O(\lambda^2)]/L}+\int_0^\infty dE^\prime
{\rho_1(E^\prime)\over E^\prime -E}$$ 
In contrast nonplanar  corrections, suppressed by powers of $1/N$,
show no such gap: 
$$R_{\rm nonplanar}(E,L)\sim {1\over N^2}\int_{-\pi\lambda/2L}^\infty dE^\prime
{\rho_2(E^\prime)\over E^\prime -E}$$ 
Thus at $N=\infty$ and arbitrarily weak coupling, 
there is a gap in the $Q{\bar Q}$ system \cite{klebanovmt}.

At strong coupling, $\lambda\gg1$, the AdS/CFT correspondence gives the
excitation spectrum of the $Q{\bar Q}$ system
in ${\cal N}=4$ supersymmetric Yang-Mills
theory as the semi-classical quantization of a IIB superstring
connecting the two sources on the boundary of AdS$_5$.
The lightcone action for the worldsheet fields representing the
coordinates of AdS$_5$ is \cite{metsaevtt}
\bea
\hskip-10pt S^{{\rm AdS}_5}_{ws}&=&\int d\tau\int_0^{p^+}d\sigma{\gamma^2\over2}
\left[\dot{\bf x}^2-
e^{2\phi}{\bf x}^{\prime2}+e^{-\phi}\dot{\phi}^2
-e^\phi{\phi}^{\prime2}\right] \nonumber\\
&&\quad 4\gamma^2\ =\ R^2T_0\ =\ \sqrt{\alpha_s N_c/\pi}\ 
=\ \sqrt{\lambda}\nonumber\label{action}
\eea  
\begin{figure}[ht]
\begin{center}
\psfrag{'R3'}{$R^3$}
\psfrag{'1/r'}{${e^{-\phi}}$}
\psfrag{'L'}{$L$}
\includegraphics[width=3in]{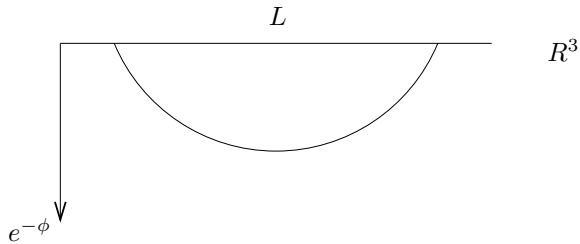}
\caption{A static string with ends on $R^3$, the boundary of AdS$_5$,
bends away from the boundary into the bulk of AdS$_5$.}
\label{static}
\end{center}
\end{figure}
The static solution (see Fig.~\ref{static}) has energy 
$E_0$ which increases as $-c\sqrt{\lambda}/L$ with separation, showing the absence
of a confining force \cite{maldacenaqqbar}.
Nonetheless, this stretched string has
an infinite number of stringy excitations.

Semi-classical quantization of the small oscillations
about this static solution \cite{callang}
gives string-like modes with discrete levels just above $E_0$:
\bea E_{N_n}-E_0 = \sum N_n\omega_n ;\qquad
\omega_n = {(2\pi)^{3/2}\over \Gamma(1/4)^2 L}\xi_n;
\qquad \xi_n\sqrt{\xi_n^4-1}\int_0^1{t^2dt\over[1+\xi_n^2t^2]
\sqrt{1-t^4}}={n\pi\over2} \nonumber
\eea
Where $n=1,2,\dots$. For large $n$, 
$\omega_n\sim (2\pi)^{3}(n+1)/\left(\Gamma(1/4)^4 L\right)$ \cite{browertt},
typical of normal modes
of a string with effective tension $T_{eff}\sim 1/L^2$.
We see that ${\cal N}=4$ is teetering on the brink
of quark confinement:
a stringy object is there for $L$ finite, 
but a mechanism to prevent $T_{eff}$
from dropping to 0 when $L\to\infty$ is lacking.
Finally, we note that near threshold ($E=0$) discrete levels 
accumulate \cite{klebanovmt}
${E_{n+1}/E_n}\sim e^{-\pi/\sqrt{4\lambda}}$. 
This is shown by a semiclassical quantization of the motion of the
midpoint of string stretched a distance $D>>L$
from the boundary of AdS, where the string ends reside. 

\begin{figure}[ht]
\psfrag{'ES'}{$\hskip-.25in-{c\sqrt{\lambda}\over L}$}
\psfrag{'EW'}{$\hskip-.25in-{\pi{\lambda}\over L}$}
\psfrag{'lg1'}{$\lambda\gg1$}
\psfrag{'ll1'}{$\lambda\ll1$}
\psfrag{'E'}{$E$}
\begin{center}
\includegraphics[width=3in]{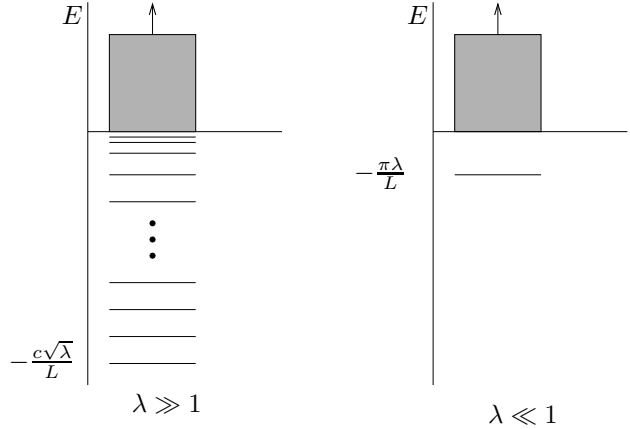}
\caption{Energy spectrum of the ${\cal N}=4$ $Q{\bar Q}$ system 
for strong and weak coupling at $N = \infty$.}
\label{levels}
\end{center}
\end{figure}

To summarize we show the  energy level diagram for the ${\cal N}=4$
$Q{\bar Q}$ system at $N = \infty$ 
for weak and strong 't Hooft coupling (see Fig.~\ref{levels}). 
The transition from weak to strong coupling is rather mundane: a
stronger binding with a deepening gap
that eventually supports excited discrete levels that peel
off the continuum and move into the gap.
$N=\infty$ is essential here: otherwise the
continuum goes all the way down to $E_0$.
At large finite $N$ this continuum 
would be dominated by very narrow resonances.

Surprisingly a simple Feynman gauge ladder diagram model
\cite{ericksonssz} shows qualitatively similar physics. 
The authors of this paper show that the weak coupling energy is recovered and 
at strong coupling the ground energy
$\propto-\sqrt{\lambda}/L$ with a different numerical coefficient than
the AdS string.
At intermediate coupling \cite{browertt} this model
shows no more bound states for $\lambda<1/4$.
For $\lambda>1/4$ an infinite number of bound states appear
with threshold behavior 
${E_{n+1}/E_n}\sim e^{-\pi/\sqrt{4\lambda-1}}$,
{\it with the same strong coupling behavior as ${\cal N}=4$}.
The big qualitative difference from the exact strong coupling 
behavior is that the ladder model shows
only a single mode of small oscillations instead
of an infinite number of stringy modes $\omega_n$. This defect
is presumably due to the neglect of all the non-ladder planar diagrams.
An accurate treatment of the ${\cal N}=4$ theory at 
intermediate coupling requires either
the proper quantization and solution of IIB superstring on AdS$_5\times$S$_5$,
or a way to sum {\it all} planar diagrams. For QCD, the absence of
a sensible strong coupling limit has hindered the discovery of its string
theory dual. 
Understanding the ${\cal N}=4$ theory by directly 
summing planar diagrams could
teach us how to do the same with real QCD at $N=\infty$.
In particular, the 
lightcone worldsheet formalism was developed as a way 
to read off the string
dual directly from the sum of planar diagrams \cite{bardakcit}.
This program also gives a strategy for implementing 
a gluon chain model of the flux tube \cite{greensitet}.

\end{document}